\documentclass[10pt,twoside]{article}
\pdfoutput=1 

\usepackage{Latex-document}
\almvol{00}
\almttone{Frontiers of Science Awards}
\almtttwo{for Math/TCIS/Phys}
\firstpage{1}
\usepackage[english]{babel}
\usepackage[utf8]{inputenc}

\usepackage{multicol}

\usepackage{tabu}
\usepackage[normalem]{ulem}

\usepackage{amsmath,xcolor}
\usepackage{amssymb,mathrsfs}
\usepackage{cite}

\RequirePackage{braket,slashed}
\RequirePackage{graphicx}

\numberwithin{equation}{section}

\definecolor{darkgreen}{rgb}{0,0.4,0}

\newcommand{\refmod}[1]{ #1}

\newcommand\mathd{\mathrm{d}} 
\newcommand\eph{\ensuremath{e_{\mathrm{ph}}}}

\newcommand\aph{\ensuremath{\alpha_{\mathrm{ph}}}}

\newcommand{\GeV}{\;\mathrm{GeV}}

\newcommand{\as}{\alpha_s}
\newcommand{\MSbar}{\ensuremath{\overline{\text{MS}}}}

\newcommand\xbj{x_{\scriptscriptstyle\rm bj}}
\newcommand\mpr{m_{\rm p}}

\newcommand\nn{ \nonumber \\ }

\newcommand\mub{ \mu}

\begin{document}

\markboth{\hfill{\rm Aneesh Manohar, Paolo Nason, Gavin Salam, and Giulia Zanderighi} \hfill}{\hfill {\rm The Photon PDF \hfill}}

\title{The photon parton distribution function: updates and applications}

\author{Aneesh Manohar, Paolo Nason, Gavin Salam, and Giulia Zanderighi\footnote{Speaker at the 2024 International Conference of Basic Science, presenting the paper \cite{Manohar:2016nzj}, which received the 2024 Frontier of Science Award in Theoretical Physics.}}

\begin{abstract}
  The photon parton distribution function (PDF) of the proton is
  crucial for precise comparisons of LHC cross sections with
  theoretical predictions. However, it was previously affected by very
  large uncertainties of around ${\cal O}(100\%)$ or dependent upon
  phenomenologically inspired models. In the
  paper~\cite{Manohar:2016nzj}, we demonstrated
  how the photon PDF could be determined using the proton structure
  functions $F_2$ and $F_L$ measured in electron--proton scattering
  experiments. We provided an explicit formula for the PDF, which can
  be systematically improved order by order in perturbation theory. We
  obtained a photon PDF with errors $\lesssim 2$\% for $10^{-4} < x <
  0.1$. Here, we recall the underlying idea and method used to obtain
  this result, as well as the progress made since then.
\end{abstract}

\maketitle


\section{Introduction}

Hard scattering processes involving hadrons can be computed in terms
of parton distribution functions (PDFs) $f_{a/H}(x,\mu^2)$, which
represent the probability of finding a parton $a$ with momentum
fraction $x$ in a hadron target $H$.\footnote{From now on, we will
  only discuss proton targets, and omit the subscript $H$.} These
distributions depend logarithmically on the renormalisation and
factorisation scale $\mu$ due to radiative corrections. This principle
underlies most studies conducted at CERN's Large Hadron Collider
(LHC). Of course, the primary partons are quarks and gluons.

It has been known for a long time that ultra-relativistic charged
particles generate an electromagnetic field that can be interpreted as
a distribution of photons, as originally calculated by Fermi,
Weizsäcker and Williams
\cite{Fermi:1924tc,vonWeizsacker:1934nji,Williams:1934ad} for
point-like particles. The same principle applies to the proton and its
constituents.
Due to the weakness of the electromagnetic coupling, the photon
density in the proton had long been considered of secondary
importance.
However, between 2010 and 2016, it was realised that knowledge of the
photon PDF was becoming more important as the measurements at the LHC
became increasingly precise. The uncertainty on the photon PDF was
limiting our ability to predict certain key reactions at the
LHC. Notable examples include the production of the Higgs boson
through $W/Z$ fusion~\cite{Ciccolini:2007ec} or in association with an
outgoing weak boson~\cite{Denner:2011id}. For $W^\pm H$ production, it
was the largest source of uncertainty~\cite{deFlorian:2016spz}. The
photon distribution was also relevant for the production of lepton
pairs~\cite{Aad:2015auj,Aad:2016zzw,Accomando:2016tah,Alioli:2016fum,Bourilkov:2016qum},
top quarks~\cite{Pagani:2016caq}, pairs of weak
bosons~\cite{Luszczak:2014mta,Denner:2015fca,Ababekri:2016kkj,Biedermann:2016yvs,Biedermann:2016guo,Yong:2016njr,Kallweit:2017khh}
and generally for electroweak corrections in almost any LHC process.
At that time, the available results on the photon PDF, such as
{MRST2004qed}~\cite{Martin:2004dh}, {NNPDF23\_qed}~\cite{Ball:2013hta}
CT14qed\_inc~\cite{Schmidt:2015zda} and the HKR16
set~\cite{Harland-Lang:2016kog}, either had large uncertainties, or
relied upon phenomenologically inspired models.

In 2015, the lack of a precise knowledge of the photon PDF became an
even more concerning issue with the observation of a 750~GeV resonance
in $\gamma \gamma$ final states by the
ATLAS~\cite{ATLAS-CONF-2015-081} and CMS~\cite{CMS-PAS-EXO-15-004}
experiments.\footnote{The resonance disappeared in 2016 when new data
  became available.}  This observation, which generated much
excitement in the high-energy physics community, suggested that
the resonance could  have been produced in $\gamma\gamma$
annihilation, highlighting the need for a sound knowledge of the
photon PDF.

In \cite{Manohar:2016nzj,Manohar:2017eqh} we derived a formula for the
photon PDF given in terms of the electromagnetic form factors and
structure functions of the proton, which are very accurately measured.
The main idea leading to this work was that in a reaction involving a
virtual, spacelike photon, one cannot really distinguish whether the
photon is incoming or outgoing. Hence, it should be possible to
express the photon PDF, typically associated with photon emission from
the proton, in terms of the proton structure function, associated with
photon absorption by the proton.  In fact, in the frame where the
proton is at rest, the proton absorbs the spacelike photon emitted by
the electron, while in the centre-of-mass frame of the lepton, the
opposite occurs.

To compute the photon PDF, we considered a simple photon-initiated
processes, and computed it in two different ways: in terms of an
(unknown) photon PDF, and in terms of the lepton-hadron scattering
hadronic tensor and form factors. Equating the two results allowed us
to extract the unknown photon PDF in terms of electron-proton
scattering data.\footnote{ This crucial observation in
  Ref.~\cite{Manohar:2016nzj} was partly inspired by Drees and
  Zeppenfeld's study of supersymmetric particle production at $ep$
  colliders~\cite{Drees:1988pp}. Our perspective was also implicit in
  Refs.~\cite{Anlauf:1991wr,Mukherjee:2003yh}.}
The precision of our result for the photon PDF was thus a consequence
of the precision of the available lepton-proton scattering data. We
stress that this data is needed even at relatively low energies,
i.e. the structure functions in the deep-inelastic (DIS) limit are not
sufficient for our purposes.  However, the required data was
available, having been accumulated over many years of nuclear physics
experiments.

In \cite{Manohar:2016nzj} we carried out the calculation using the
production of a hypothetical heavy lepton in electron-proton
collisions, mediated by a flavour violating magnetic moment
interaction, as a hard probe.  We provided a formula for the photon
PDF of the proton $f_\gamma(x,\mu^2)$ as an integral over proton
structure functions $F_2(x,Q^2)$ and $F_L(x,Q^2)$, including all terms
of order $\mathcal{O}(\alpha)$ and with errors of order
$\mathcal{O}(\alpha^2L)$ and $\mathcal{O}(\alpha\alpha_s)$, where $L$
is the logarithm of $\mu$ divided by a typical hadronic
scale.\footnote{The photon PDF $f_\gamma(x,\mu^2)$ is of order $\alpha
  L$. The result in Ref.~\cite{Manohar:2016nzj} included terms of
  order $\alpha^2 L^2 (\as L)^n$ and $\alpha (\alpha_s L)^n$, but not
  of order $\alpha^2 L$.  By assuming $L\approx 1/\alpha_s$, and
  considering that $\alpha_s^2 \approx \alpha$, we see that the first
  subleading terms to include are of order $\alpha$ (i.e. one power of
  $L$ less than the leading term) and of order $\alpha^2 L^2$.}
This formula enabled us to obtain the photon PDF with an uncertainty
of less than $3\%$ over a wide range of $x$ values, $10^{-5} \le x \le
0.5$. This reduced the uncertainty by a factor of about forty compared
to previous photon PDF determinations, such as
{MRST2004qed}~\cite{Martin:2004dh} and
{NNPDF23\_qed}~\cite{Ball:2013hta}, which relied on fits to LHC data
and/or modelling.

The derivation in \cite{Manohar:2016nzj} was extended in
\cite{Manohar:2017eqh}, where we re-derived our formula for the
photon PDF using as a hard probe a hypothetical scalar produced via
$\gamma\gamma$ annihilation in proton-photon collisions, thus checking
the universality of our result.  Furthermore, we provided another
independent derivation by defining the photon PDF as the light-cone
Fourier transform of the two-point function of electromagnetic
field-strength tensors. This approach gives a representation for the
photon PDF that is \emph{exact}, i.e. valid to all orders in the
strong and electromagnetic interactions, and that can in principle be
used to compute the photon PDF at arbitrarily high orders in
perturbation theory.  It was used to obtain expressions for the photon
PDF including all terms of order $\alpha \alpha_s$ and $\alpha^2$, one
order higher in $\alpha_s(\mu)$ or $\alpha(\mu)$ than the result in
\cite{Manohar:2016nzj}. Furthermore, results were also given for
polarised photons and transverse-momentum dependent PDFs.

\section{Definitions and notation}\label{sec:em}

We define the physical electromagnetic $\eph$ in terms of the vacuum
polarisation function $\Pi(q^2,\mub^2)$,
\begin{align}
\eph^2(q^2) &= \frac{e^2(\mub^2)}{1 - \Pi(q^2,\mub^2)}\,,
\label{3.15}
\end{align}
which depends on $q^2$, but is independent of $\mub^2$. Thus $\aph(0)
= \eph^2(0)/4\pi$ is the usual fine structure constant $\simeq
1/137.036$ measured in atomic physics. Eq.~(\ref{3.15}) is used in the
spacelike region $q^2<0$, where $\Pi(q^2,\mub^2)$ is real.

The proton hadronic tensor, which enters the cross section formula for
deep-inelastic scattering, is defined as
\begin{align}
W_{\mu\nu}(p,q,s) = \frac{1}{4\pi}\int \mathd^4 z \ e^{i q \cdot z}
\langle p,s| \left[ j_\mu(z), j_\nu(0)\right] | p,s\rangle \,,
\label{eq:hadtens}
\end{align}
where $j_\mu$ is the electromagnetic current, $p$ is the incoming
proton momentum, $q$ is the photon momentum (transferred to the
proton) and $s$ denotes the proton spin.
The conventional decomposition of $W_{\mu \nu}$ into structure
functions is
\begin{align}
W_{\mu\nu}(p,q,s) &= F_1 \left(-g_{\mu\nu} + {q_\mu
q_\nu\over q^2}\right) + {F_2\over p \cdot q} \left(p_\mu - {p\cdot q \ q_\mu\over
q^2}\right)
\left(p_\nu - {p\cdot q\ q_\nu\over q^2}\right)\cr
\noalign{\smallskip}&+ {ig_1\over p\cdot q}\ \epsilon_{\mu\nu\lambda\sigma}
q^\lambda s^\sigma +
{ig_2\over (p\cdot q)^2}\ \epsilon_{\mu\nu\lambda\sigma}
q^\lambda \left( p\cdot q
\, s^\sigma - s\cdot q\,p^\sigma\right)\ .
\label{3.21}
\end{align}
The proton spin four-vector $s$ is normalised so that $p \cdot s=0$,
$s \cdot s = - \mpr^2$, where $\mpr$ is the proton mass. The structure
functions $F_1$, $F_2$, $g_1$ and $g_2$ are functions of
\begin{align}
\xbj &= \frac{Q^2}{2 p \cdot q}
\label{3.22}
\end{align}
and $Q^2$. We also introduce the longitudinal structure function
\begin{align}
  F_L(\xbj,Q^2) &\equiv \left(1+\frac{4\xbj^2\mpr^2}{Q^2}\right)F_2(\xbj,Q^2)  - 2\xbj F_1(\xbj,Q^2), 
 \label{3.28}
\end{align}
and write our results using $F_2$ and $F_L$ instead of $F_2$ and
$F_1$.

\section{Photon PDF Formula}\label{sec:probe}

The formula for the photon PDF was obtained by using as a hard probe a
toy-process, the scattering of a heavy neutral lepton off a proton via
a magnetic moment interaction. We verified the result by carrying out
the same derivation using as a probe the production of a scalar
particle via photon-photon fusion and also by using the definition of
PDF operators in terms of light-cone Fourier transforms
in~\cite{Manohar:2017eqh}.

It is convenient to split the photon PDF into two pieces,
\begin{align}
    \label{eq:fgamma-split}
    f_\gamma(x,\mub^2) &= f^{\text{PF}}_\gamma(x,\mub^2) +
                         f^{\text{con}}_\gamma(x,\mub^2) \,,
\end{align}
referred to as the ``physical factorisation'' and
``$\MSbar$-conversion'' terms.  The $\MSbar$-conversion is an
additional finite contribution to the photon PDF obtained by applying
the $\MSbar$ subtraction scheme to regulate UV divergences in the
factorization formula order-by-order in perturbation theory.

The physical factorization term is exact,
\begin{align}
      \label{6.12a}      
    f^{\text{PF}}_\gamma(x,\mub^2)
    &= \frac{1 }{2 \pi \alpha(\mub^2)  x} 
      \int_x^1 \frac{\mathd z}{z}  \int_{\frac{m_p^2 x^2}{1-z}}^{\frac{\mub^2}{1-z}}  \frac{\mathd Q^2}{Q^2}  \aph^2(\refmod{-Q^2})
      \nn
    & \biggl\{- z^2 F_L(x/z,Q^2) + \left[ 2-2z  +  z^2  + \frac{2
      m_p^2 x^2}{Q^2}\right] F_2(x/z,Q^2)\biggr\}\,.
\end{align}
The $\MSbar$-conversion term has the perturbative expansion
\begin{align}
    \label{9.11b}
     f^{\text{con}}_\gamma(x,\mub^2) 
    &= \sum_{r \ge 0,s \ge 1} \left( \frac{\alpha_s}{2\pi}\right)^r 
      \left( \frac{\alpha}{2\pi}\right)^s  f^{\substack{(r,s)  }}_\gamma(x,\mub^2)\,,
  \end{align}
where the couplings are evaluated at $\mu^2$.
At leading order the $\MSbar$ -conversion term is~\cite{Manohar:2016nzj} 
\begin{align}
 f^{\substack{(0,1)  }}_\gamma(x,\mub^2)
& = \frac{1}{x}
 \int_x^1 \frac{\mathd z}{z}  (   -  z^2 )   F_2 (x/z,\mub^2)
  +\mathcal{O}(\alpha^2,\alpha\alpha_s) \,. \label{6.15a}
\end{align}
The next order contributions $f_\gamma^{(1,1)}$
and $f_\gamma^{(0,2)}$ are given in~\cite{Manohar:2017eqh}.

\section{Experimental Inputs}
\label{sec:input-data}

To evaluate the photon parton density we require information on the
$F_2$ and $F_L$ structure functions over the full $x,Q^2$ allowed
kinematic range.
Using $L \sim \ln \mu^2/m_p^2$, we include terms $\alpha L (\as L)^n$
at lowest order and $\alpha (\as L)^n$, $\alpha^2 L^2 (\as L)^n$
corrections at higher order. The physical factorization term is
enhanced by $L$ relative to the $\MSbar$-conversion term due to the
additional integral over $Q^2$ in Eq.~\eqref{6.12a}.

The $F_2$ and $F_L$ structure functions are most commonly determined
from electron--proton scattering data.
We separate the data inputs according to the kinematic region and the
corresponding final state in $ep$ scattering.
The main kinematic variables for the separation will be $Q^2$ and
$W^2$ where
\begin{equation}
  \label{eq:Wsq}
  W^2 = m_p^2 + \frac{1-\xbj}{\xbj} Q^2 \,, 
\end{equation}
is the squared invariant mass of the outgoing system associated with
the hadronic side of the collision.

\noindent {\bf Elastic contribution:}
In our definition, the elastic contribution corresponds to the region
of $W < \mpr + m_{\pi^0}$.
In particular it includes configurations where one or more photons are
radiated from the proton.%
\footnote{For the determination of the
  structure functions we find it useful to think of a process in which
  there can be at most one exchanged photon between the probe and the
  proton, as in the process of Sec.~\ref{sec:probe}.
  In actual electron-proton scattering experiments there can be two or
  more exchanged photons, either real of virtual.
  These corrections are beyond our accuracy and cannot be classified
  in terms of the usual electromagnetic structure functions, since
  they correspond to a more complex tensor structure.}
Experimental data on elastic scattering is usually corrected for
radiation from the proton, since the measurements are performed with
the goal of extracting the electric and magnetic Sachs form factors of
the proton, $G_E$ and $G_M$ respectively.
A widely used approximation for the $G_{E,M}$ form factors is the
dipole form,
\begin{equation}
  \label{eq:dipole-form-factor}
  G_E^\text{dip}(Q^2) = \frac{1}{(1+Q^2/m_\text{dip}^2)^{2}} \,,\qquad
  G_M^\text{dip}(Q^2) = \mu_p G_E(Q^2)\,,
\end{equation}
where $m_\text{dip}^2=0.71\GeV^2$ and $\mu_p\simeq 2.793$ is the
anomalous magnetic moment of the proton.
For $Q^2 = 0$ this form yields the exact results $G_E(0) = 1$ and
$G_M(0)=\mu_p$, while elsewhere it is an approximation.

\begin{figure*}
  \centering
  \includegraphics[page=1,width=0.5\textwidth]{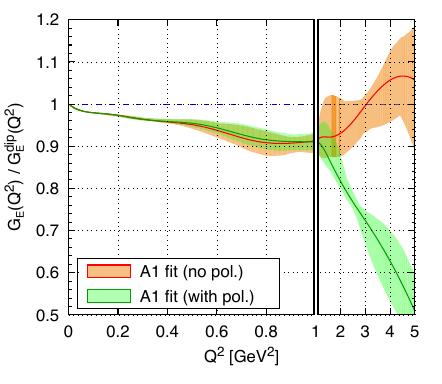}%
  \hfill%
  \includegraphics[width=0.5\textwidth]{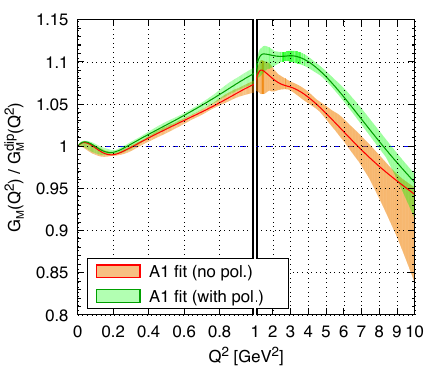}%
  \caption{Elastic form factors (ratio to standard dipole form) for
    the electric (left) and magnetic (right) as fitted by the A1
    collaboration~\cite{Bernauer:2013tpr} with and without polarised
    data. 
    Note the change in scale at $Q^2=1\GeV^2$ along the $x$ axis.
  }
  \label{fig:elastic-FF}
\end{figure*}

For accurate results, the dipole approximation,
Eq.~(\ref{eq:dipole-form-factor}), is not sufficient.
The most recent extensive experimental study of the form factors comes
from the A1 collaboration~\cite{Bernauer:2013tpr}.
The A1 data itself is limited to $Q^2 \lesssim 1\GeV^2$, however the
work includes fits to global data up to $Q^2 \sim 10\GeV^2$.
The electric and magnetic form factor fits are shown in
Fig.~\ref{fig:elastic-FF}, normalised to the dipole form.
Both fits show clear deviations from the dipole form. 
Note that the fits extend only up to $Q^2 = 10\GeV^2$ and beyond this
point we use the dipole shape, normalised to the fitted $G_{E/M}(Q^2)$
at $Q^2=10\GeV^2$. 
We treat the fit uncertainties on the elastic and magnetic components
as $100\%$ correlated, which is the most conservative assumption. 
%

\noindent {\bf Low-$Q^2$ region:}
In addition to the elastic component, we need the inelastic part.
This corresponds to the region of $W > \mpr + m_{\pi^0}$.
We split the inelastic part into several sub-regions.
At low $Q^2$, the structure functions cannot be computed
from parton distribution functions and we must
rely on direct measurements and theoretical model-independent
constraints.

\begin{figure}[h]
  \centering
  \includegraphics[width=0.55\textwidth]{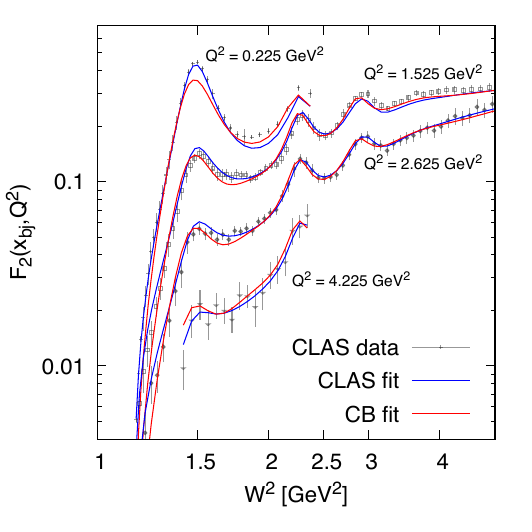}\hfill%
  \includegraphics[width=0.44\textwidth]{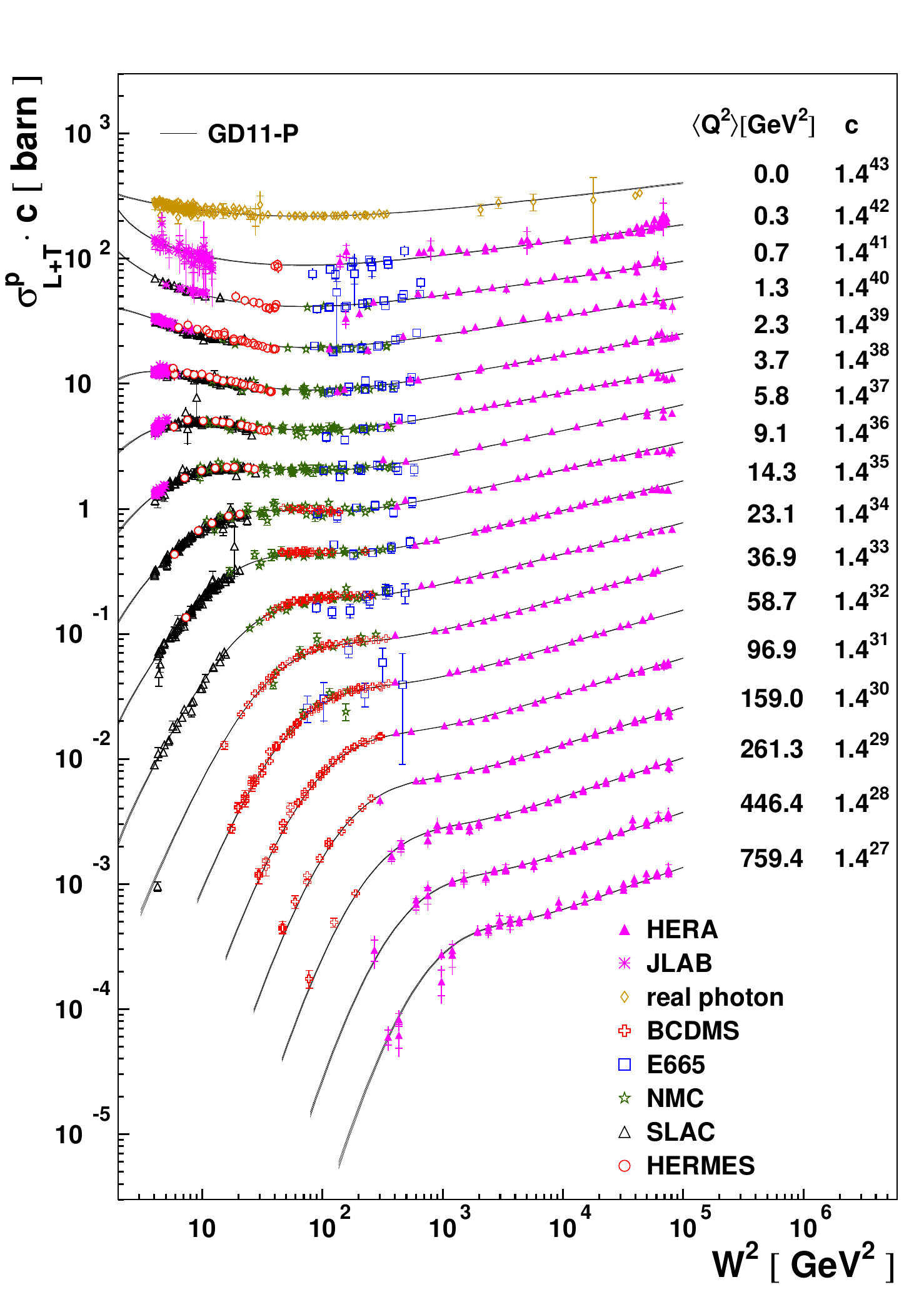}
  \caption{Left: illustration of a subset of the CLAS
    data~\cite{Osipenko:2003bu} in the resonance region, compared to
    two fits, one from the CLAS paper and the other from Christy and
    Bosted (CB)~\cite{Christy:2007ve}.
    The CLAS dataset covers the range $0.225 \le Q^2/\text{GeV}^2 \le 4.725$ in
    steps of $0.05$, with a quality comparable to that shown in the plot across
    the whole $Q^2$ range.
    The errors on the data correspond to the sum in quadrature of
    statistical and systematic components.
    The CLAS data is only a small part of the data that is available
    in the resonance region and used for the fits (see  also Fig.~6 of
    Ref.~\cite{Osipenko:2003bu}).  
    Right: illustration of the GD11-P fit from the HERMES
    collaboration~\cite{Airapetian:2011nu} and corresponding data in
    the continuum region
    (plot reproduced with kind permission of the HERMES
    collaboration). 
}
  \label{fig:clas-data-v-fits}
\end{figure}

There is a wealth of data covering the low $Q^2$ region. 
Rather than using these data directly, we will rely on existing fits
of those data.
Fits generally focus upon either the resonance region, $W^2 \lesssim
3\GeV^2$ or the continuum region, $W^2 \gtrsim 4\GeV^2$.
Fig.~\ref{fig:clas-data-v-fits} (left) shows data from the CLAS
experiment~\cite{Osipenko:2003bu}, compared to two global fits to
resonance region data.
The figure (which includes only a small subset of the available data)
illustrates the coverage in $Q^2$ and the quality of the available
data.
The data is shown as a function of $W^2$ in order to clearly
show the resonance peaks starting with the $\Delta$  resonance and beyond.
The CLAS fit is intended for use only for $Q^2 > 0.5\GeV^2$, while
the Christy-Bosted~\cite{Christy:2007ve} fit is intended for use down
to $Q^2 = 0$ and explicitly includes photoproduction data.
Comparing the CLAS and Christy-Bosted fits at $Q^2$ values below the
quoted validity range of the CLAS fit shows, however, that they are
relatively similar.

For the continuum region, the HERMES collaboration has provided a fit,
GD11-P~\cite{Airapetian:2011nu}, using a wide range of data and the
ALLM~\cite{Abramowicz:1991xz} functional form.
Fig.~\ref{fig:clas-data-v-fits} (right), taken from
Ref.~\cite{Airapetian:2011nu}, illustrates the good quality of the
fit. 
Careful inspection of the figure reveals that at each $Q^2$ value the
fit consists of three lines, whose separation represents the
uncertainty.

%
\noindent {\bf High-$Q^2$ continuum:}
For sufficiently large $Q^2$ and $W^2$ one can calculate $F_2$ and
$F_L$ from parton distribution functions (PDFs) using the known
perturbative expansion of the DIS coefficient functions.
This is more reliable than using a fit to available data (e.g.\ GD11-P
also includes some high-$Q^2$ data), because the extension to
arbitrarily large $Q^2$ is provided by DGLAP evolution rather than the
extrapolation provided by some a priori arbitrary parametrisation.
Furthermore, in recent years there has been extensive progress in the
extraction of PDFs from DIS and collider data, including detailed and
well-tested uncertainty estimates.

\begin{figure}
  \centering
  \includegraphics[width=0.49\textwidth,page=1]{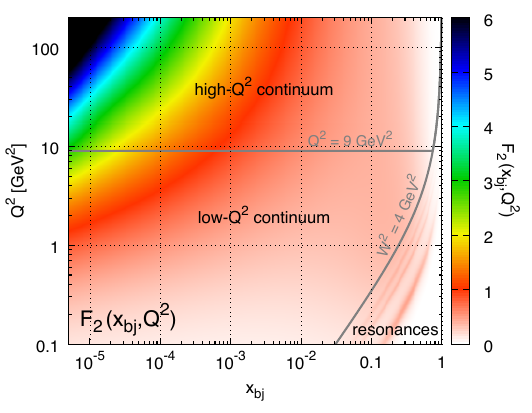}%
  \hfill
  \includegraphics[width=0.49\textwidth,page=2]{Figures/F2-plane.pdf}
  \caption{Values of the structure functions $F_2$ (left) and $F_L$
    (right) as a
    function of $\xbj$ and $Q^2$, using a PDF4LHC15\_nnlo\_100-based
    NNLO ZM-VFNS prescription for $Q^2>9\GeV$ and $W^2>4\GeV^2$, and
    the CLAS+GD11-P combination elsewhere.}
  \label{fig:F2FL-plane-full}
\end{figure}

Our default domain for using a PDF-based evaluation of the structure
functions will be $Q^2 > Q^2_\text{PDF}  = 9\GeV^2$ and $W^2 > 4\GeV^2$.
In the rest of the kinematic plane we will use the resonance and low
$Q^2$ continuum fits.
The breakup of the $\xbj{-}Q^2$ plane is summarised in
Fig.~\ref{fig:F2FL-plane-full}.
The colour-coding provides a visualisation of the size of the
integrand in the sum of Eqs.~(\ref{6.12a}\,,\,\ref{9.11b}).

\section{Numerical Results and Uncertainties}
\label{sec:uncertainties}

\begin{figure}
 \includegraphics[width=0.48\textwidth,page=1]{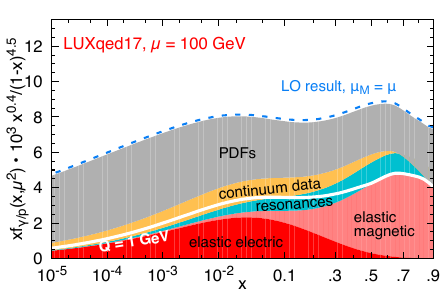}\hfill
  \includegraphics[width=0.48\textwidth,page=8]{Figures/contribs.pdf}
  \caption{The left panel shows the contributions to the photon PDF at
    $\mu=100$~GeV,
    multiplied by $10^3 x^{0.4}/(1-x)^{4.5}$, with a breakdown into the various
    components discussed in the text. 
    The white line is the sum of the inelastic contribution from $Q^2 \le 1\,
    \text{(GeV)}^2$ in Eq.~(\ref{6.12a}) and the full
    elastic contribution. 
    The full physical factorisation result of Eq.~(\ref{6.12a}), which is equivalent to a
    LO result, is given by the
    dashed blue line. 
    The right panel shows the same plot for $\mu=500$\,GeV, with the
    scale $\mu_M$ for the LO results
    set to $\mu/2$ or $2\mu$. 
    The total PDF (edge of grey region) is shown for $\mu_M=\mu$. The
    $\MSbar$-conversion term (difference between grey region and
    dashed blue curve) 
    has a significant impact with scale choices other than $\mu_M=\mu$.
}
  \label{fig:breakup}
\end{figure}

The photon PDF computed using the method of this paper is given in the
left-hand panel of Fig.~\ref{fig:breakup} for $\mu=100$\,GeV.
It is rescaled by a factor $10^3 x^{0.4}/(1-x)^{4.5}$ to facilitate
the simultaneous study of different $x$ regions.
The plot includes a breakup into the different contributions discussed in
Sec.~\ref{sec:input-data}. 
There is a sizeable elastic contribution, with an important magnetic
component at large values of $x$.
The resonance and continuum regions are also quantitatively relevant.
The white line represents contributions arising from the $Q^2<1$
region of all the structure functions, including the full elastic
contribution, and this serves to illustrate the importance of a proper
inclusion of the low $Q^2$ region, given the accuracy we aim for.
The PDF contribution, in the physical factorisation scheme, is from
the bottom of the grey region to the blue dashed curve.
The $\MSbar$-conversion term, Eqs.~(\ref{9.11b}), is negative and
corresponds to the difference between the blue dashed curve (PF
result) and the top edge of the grey region (final full result in the
$\MSbar$ scheme).

The right-hand plot of Fig.~\ref{fig:breakup} illustrates how the
components evolve when increasing the factorisation scale $\mu$.
The main change is associated with the $\log Q^2$ growth of the ``PDF''
contribution and is most important at small $x$ values, a consequence
of the fact that the quarks distributions themselves increase rapidly
with $Q^2$ at small $x$.
The elastic, resonance and (low-$Q^2$) continuum contributions to the
photon PDF all depend slowly on $\mu$ via the overall
$1/\alpha(\mu^2)$ factor in Eq.~(\ref{6.12a}).
These components, though formally NLO, remain a significant fraction
of the overall photon PDF, even at this large value of $\mu$.

The right panel in Fig.~\ref{fig:breakup} also shows the impact of
scale variation on the contributions at $\mu=500$\,GeV. 
The blue dashed curves are for $\mu_M=\mu/2$ and
$\mu_M=2\mu$.\footnote{The scale $\mu_M$ is used instead of $\mu$ to
  split the photon PDF into physical factorization and $\MSbar$
  conversion terms. See~\cite{Manohar:2017eqh} for details.}
The total $\MSbar$ photon PDF (the top edge of the grey band) uses our
central choice of $\mu_M=\mu$.
The impact 
of a change of $\mu_M$ on the total photon PDF would be
barely visible in the plot, because the substantial scale dependence
of the PF result is largely cancelled by that of the
$\MSbar$-conversion term.
Previous photon PDFs were at best accurate to leading order, and hence
had much larger scale uncertainties than our determination, which we
dubbed ``LUXqed''.

\begin{figure}
  \centering
  \includegraphics[width=0.6\textwidth,page=2]{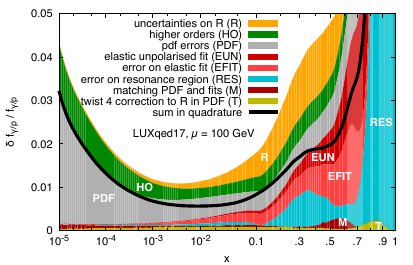}
  \caption{Breakdown of uncertainties on the photon distribution. 
    The uncertainties are shown stacked linearly, while the sum in
    quadrature, i.e.\ our final uncertainty, is represented by the
    thick black line. See the text for a detailed description of the
    various contributions.
  }
  \label{fig:uncertainty-breakdown}
\end{figure}

The final uncertainty on our PDF distribution is taken to be the sum
in quadrature of many individual uncertainty sources, because they are
uncorrelated. The individual contributions are discussed in detail in
\cite{Manohar:2017eqh}. The various uncertainties with labels as in
Fig.~\ref{fig:uncertainty-breakdown} are:
\begin{itemize}
\item[(EFIT)] The uncertainty on the elastic contribution that comes
  from the uncertainty on the A1 world polarised form factor fits.
  This band is asymmetric and we symmetrise it using the largest
  deviation to obtain a (more conservative) symmetric band.

\item[(EUN)] The uncertainty that comes from replacing the A1 world
  polarised fit (which includes a two-photon-exchange correction) with
  just the world unpolarised data (which does not).
  This provides a one-sided uncertainty, which we again symmetrise. 

\item[(RES)] We replace the CLAS resonance-region fit with the
  Christy-Bosted fit.
  This replacement gives a one-sided uncertainty, which we once again
  symmetrise. 

\item[(R)] A modification of $R_\text{L/T}$ (used to parameterize $F_L$) by $\pm 50\%$
  around its central value.
\item[(M)]  A modification of the $Q^2_\text{PDF}$ scale which governs the
  transition from the GD11-P structure function fit to a PDF-based
  evaluation.
  The default choice of $Q^2_\text{PDF} = 9\GeV^2$ is reduced to
  $5\GeV^2$ and since this is a one-sided uncertainty, the resulting
  effect is symmetrised.

\item[(PDF)] The input PDF uncertainties for $Q^2 > Q_\text{PDF}^2$ according
  to the default prescription for the PDF
  (PDF4LHC15\_nnlo\_100~\cite{Butterworth:2015oua}).

\item[(T)] A twist-four modification of $F_L$.  This is a one-sided
  modification that is then symmetrised.

\item[(HO)] An estimate of missing higher-order effects obtained by taking
  the largest deviation of any of the NNLO results  relative to the NLO result.
  The resulting uncertainty is symmetrised.
\end{itemize}

The impact of the different sources of uncertainty is shown in
Fig.~\ref{fig:uncertainty-breakdown}, and our final uncertainty, shown
by the black line, is given by adding the contributions in
quadrature.\footnote{There are correlations between high and low $Q^2$
  that have not been included in our analysis. For example, low $Q^2$
  values for $F_2(x,Q^2)$ are correlated with quark and gluon PDFs at
  high $Q^2$, via DGLAP evolution of $F_2$.}
The overall uncertainty is less than 2\% for $10^{-4} < x < 0.1$. For
small values of $x$, the uncertainty is dominated by the uncertainties
in the parton distributions of quarks (and gluons), which enter the
high-$Q^2$ part of the photon PDF integral. 
As $x \to 1$, the
uncertainties are dominated by the low-$Q^2$ part of the photon PDF
integral from elastic form factors, the resonance contribution, and
$\sigma_L$. 
This is a reflection of the fact that non-perturbative effects (such
as higher twist corrections) grow like
$\Lambda_\text{QCD}^2/[Q^2(1-x)]$ as $x \to 1$, and that, for $x$
close to $1$, quark PDFs fall off rapidly as $Q^2$ increases, so the
low-$Q^2$ region contributes significantly to $f_\gamma$ as $ x\to 1$.

The resulting PDF was released in LHAPDF format~\cite{Buckley:2014ana}
as sets that augment the PDF4LHC15 set, notably as
\texttt{LUXqed17\_plus\_PDF4LHC15\_nnlo\_100} and
\texttt{LUXqed17\_plus\_PDF4LHC15\_nnlo\_30}.
Note that this also involved adjustments of the distributions of other
flavours so as to ensure a consistent momentum sum, as well as the
correct QED evolution of the quark flavours.

\section{Recent developments, applications and spin-offs}
\label{sec:spinoffs}

\subsection{Global fits including the photon PDF}
\begin{figure}
  \centering
  \includegraphics[width=0.7\textwidth]{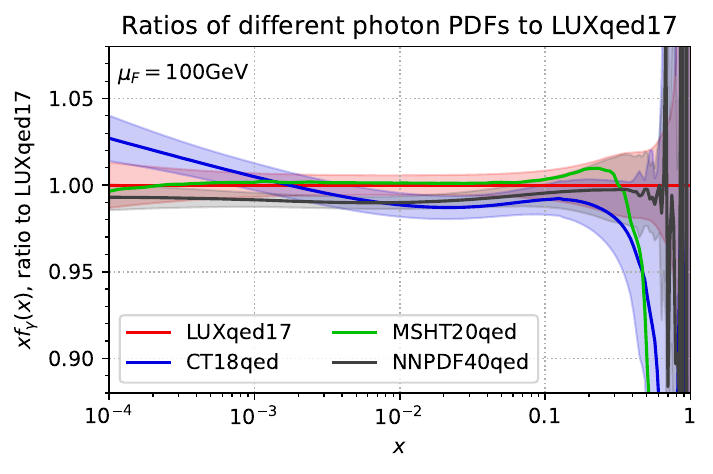}
  \caption{Comparison of recent determinations of photon distributions
    based on the LUXqed method, taking the latest NNLO set from each
    group. }
  \label{fig:recent-photons}
\end{figure}

Since the original proposal, all major PDF fitting collaborations have
adopted the LUXqed approach for the determination of the photon.
In particular the NNPDF collaboration has included it as part of the
NNPDF3.1luxQED~\cite{Bertone:2017bme} set as well as the more recent
NNPDF4.0QED~\cite{NNPDF:2024djq} and the NNPDF40MC~QED sets for Monte
Carlo event simulation \cite{Cruz-Martinez:2024cbz}.
The MSHT group adapted the approach to develop the Ad Lucem framework
\cite{Harland-Lang:2019pla}, yielding MSHT20qed~\cite{Cridge:2021pxm}
(including separation into elastic and inelastic PDFs) and
incorporating it also into their recent approximate
N3LO~\cite{McGowan:2022nag} fits
MSHT20qed\_an3lo~\cite{Cridge:2023ryv}.
The CTEQ-TEA collaboration developed the CT18qed
set~\cite{Xie:2021equ}, which apply
Eqs.~(\ref{eq:fgamma-split})--(\ref{9.11b}) across scales or at an
initial scale followed by QED evolution and they have also used the
same approach to determine the photon content of the
neutron~\cite{Xie:2023qbn}.
One advantage of carrying out a full PDF fit at the same time as the
photon-distribution determination is that it makes it possible to
consistently take into account the interplay between the photon and
other partons in the PDF.
Fig.~\ref{fig:recent-photons} compares the photon distributions from
different groups, illustrating the generally good agreement between
them.
The small residual differences that are visible may have several
origins, for example different high-scale quark distributions, and the
choice of the initial scale where Eq.~(\ref{eq:fgamma-split}) is used
to evaluate the photon PDF (LUXqed17 and NNPDF40qed use a scale of
$100\,\text{GeV}$ to minimise higher-twist effects in the photon,
whereas CT18qed and MSHT20qed use scales respectively of $1.3$ and
$1\,\text{GeV}$ respectively).

\subsection{Lepton PDFs}
Virtual leptons in the proton can only arise through the production of
an intermediate photon. It then turns out that the lepton PDFs can be
computed from the hadronic structure functions and form factors, using
the same experimental input used in the case of the photon.  The
computation (which is considerably more complex) was carried out in
ref.~\cite{Buonocore:2020nai}, and it led to the generation of proton
PDFs including photons and leptons. It was pointed out there that
lepton scattering processes were calculable in terms of lepton PDFs,
and that in some cases they could be detectable as back to back pairs
of leptons with different flavours and/or with the same electric
charge.  Furthermore, some BSM processes became calculable. In
particular, resonant lepto-quark production was a particularly
interesting one, and was studied in \cite{Buonocore:2020erb}.  The
calculation of lepton scattering processes and of lepto-quark
production were extended to next-to-leading order
in~\cite{Buonocore:2021bsf,Greljo:2020tgv,Buonocore:2022msy,Haisch:2020xjd}.

\subsection{Experimental studies using LUX photon or lepton PDFs}

The LUX photon PDF has been instrumental in precise predictions of
photon-initiated processes, as well as in the calculation of
electroweak corrections to hadronic cross sections at the LHC. This
has significantly improved the accuracy of theoretical predictions and
enhanced the reliability of comparisons with experimental data.
Accordingly, the LUX determination of the photon PDF has been widely
utilized in a variety of Standard Model measurements, which either
involve the photon PDF directly at leading order or indirectly through
higher-order processes, such as those involving electroweak (EW)
corrections.
For instance it has been used in

\begin{itemize}
\item the measurements of $t \bar t$ differential cross sections using
  events containing two leptons~\cite{CMS:2018adi},

\item the measurement of differential cross sections for the
  production of top quark pairs and of additional jets in
  lepton+jets~\cite{CMS:2018htd},

\item the measurement of top differential kinematics, including EW
  corrections~\cite{ATLAS:2019hxz},

\item the measurement of the differential Drell-Yan cross
  section~\cite{CMS:2018mdl},

\item the exclusive $\gamma \gamma \to \mu^+\mu^-$
  production~\cite{ATLAS:2017sfe},

\item the measurement of the top charged asymmetry including EW
  effects~\cite{ATLAS:2022waa},

\item the measurement of the mass dependence of the transverse
  momentum of lepton pairs in Drell-Yan production~\cite{CMS:2022ubq},

\item the measurement of the Drell-Yan forward-backward asymmetry at
  high dilepton masses in proton-proton~\cite{CMS:2022uul},

\item the measurement of the Higgs boson width and evidence of its
  off-shell contributions to $ZZ$ production~\cite{CMS:2022ley}.

\end{itemize}

Furthermore, the LUX determination of the photon and lepton PDFs has
played a critical role in a wide array of searches for new physics at
the LHC. These PDFs are essential for accurately modeling photon- and
lepton-initiated processes, which are often key components in the
search for signals beyond the Standard Model. By providing more
precise predictions for the SM backgrounds in these channels, the LUX
PDFs have improved the sensitivity of these searches, allowing for
tighter constraints on new physics scenarios. Some specific LHC
searches where the LUX PDFs have been utilized include:

\begin{itemize}
\item the search for heavy long-lived multi-charged
  particles~\cite{ATLAS:2023zxo},

\item the search for magnetic monopoles and stable particles with high
  electric charges~\cite{ATLAS:2023esy},

\item the search for heavy neutral Higgs bosons decaying into a top
  quark pair~\cite{ATLAS:2024vxm},

\item the search for vector-boson resonances decaying into a top quark
  and a bottom quark~\cite{ATLAS:2023ibb},

\item the search for heavy neutral leptons in final states with
  electrons, muons, and hadronically decaying tau
  leptons~\cite{CMS:2024xdq},

\item the search for $t$-channel scalar and vector leptoquark exchange
  in the high mass dimuon and dielectron spectrum~\cite{CMS:2024bej},

\item the search for new physics in final states with a single photon
  and missing transverse momentum~\cite{CMS:2018ffd},

\item the search for Scalar Leptoquarks Produced via
  $\tau$-Lepton–Quark Scattering~\cite{CMS:2023bdh},

\item the search for new physics in the lepton plus missing transverse
  momentum final state~\cite{CMS:2022krd},

\item the search for resonant and non-resonant new phenomena in
  high-mass dilepton final states~\cite{CMS:2021ctt},

\item the search for contact interactions and large extra dimensions
  in the dilepton mass spectra~\cite{CMS:2018nlk},

\item the search for heavy Majorana neutrinos in same-sign dilepton
  channels~\cite{CMS:2018jxx},

\item the search for a new scalar resonance decaying to a pair of Z
  bosons~\cite{CMS:2018amk},

\item the search for high-mass resonances in dilepton final
  states~\cite{CMS:2018ipm},

\item the search for heavy neutral leptons in events with three
  charged leptons~\cite{CMS:2018iaf}.
\end{itemize}

\section{Conclusions}
\label{sec:conclu}

The photon PDF is important for both precision physics and LHC
searches, as the experiments are now sensitive to electromagnetic (and
electroweak) radiative corrections. The photon PDF is a
non-perturbative quantity. Our method allows us to compute it by
dividing it into (a) a non-perturbative physical factorisation
contribution which is determined exactly in terms of measured
non-perturbative quantities, the deep-inelastic structure functions
and electric and magnetic proton form factors (b) The $\MSbar$
conversion term which can be computed relatively simply in
perturbation theory.

Our method reduced the error on the photon PDF by a factor of forty
over previous results, to $\lesssim 2$\%. The computation is a
(unique?) example where great improvement in precision for a QCD
quantity was achieved without very involved calculations. It also
provides a nice example where experiments over a wide energy range,
from low-energy nuclear form factor and $ep$ scattering measurements
to higher energy deep-inelastic scattering combine to give a result
that is important for high energy physics and BSM searches. It opens
up many new possibilities to use data from the Large Hadron Collider.

\section*{Acknowledgements}

The work done in Refs.~\cite{Manohar:2016nzj,Manohar:2017eqh} was supported in part by ERC Consolidator Grant HICCUP
(No.\ 614577), ERC Advanced Grant Higgs@LHC (No.\ 321133), a Simons Foundation grant (\#340281 to AM), by DOE grant
DE-SC0009919, and NSF grant NSF PHY11-25915.
AM and PN acknowledge CERN-TH for hospitality,  GPS and GZ thank KITP, and AM, GPS and GZ thank MIAPP for hospitality while parts of the work were
being completed.
We thank ICBS for having selected our paper~\cite{Manohar:2016nzj} for
a FSA award in theoretical physics.

\bibliographystyle{test}
\bibliography{photon}

\address{Department of Physics,  University of California at San Diego, \\
9500 Gilman Drive, La Jolla, CA 92093-0319, USA.\\
\email{amanohar@ucsd.edu}}

\address{Universit\`a di Milano-Bicocca and INFN, Sezione di Milano-Bicocca, \\
Piazza della Scienza 3, 20126 Milano, Italy.\\
\email{paolo.nason@mib.infn.it}}

\address{%
  Rudolf Peierls Centre for Theoretical Physics, \\Clarendon Laboratory,
  Parks Road, Oxford OX1 3PU, UK\\
  and
  All Souls College, Oxford OX1 4AL, UK\\
  \email{gavin.salam@physics.ox.ac.uk}
}

\address{Max-Planck-Institut f\"ur Physik, Boltzmannstrasse 8, and \\
  Technische Universit\"at M\"unchen, James-Franck-Strasse 1,\\ 
  85748 Garching, Germany.
\email{zanderi@mpp.mpg.de}}

\end{document}